\documentclass[aps,prd,twocolumn,showpacs,amsmath,amssymb,nofootinbib,eqsecnum, preprintnumbers]{revtex4-1}
\usepackage{bbm}

\usepackage{graphicx}
\usepackage[breaklinks]{hyperref}
\usepackage{booktabs}

\newcommand{\be}{\begin{equation}}

\newcommand{\ee}{\end{equation}}
\newcommand{\n}[1]{\label{#1}}
\newcommand{\BM}[1]{\boldmath{#1}}
\newcommand{\ve}{\varepsilon}

\begin{document}
\title{Black holes, hidden symmetry and complete integrability: Brief Review}
\author{Valeri P. Frolov}
\address{Theoretical Physics Institute, University of Alberta, Edmonton, Alberta T6G 2E1, Canada}

\email{vfrolov@ualberta.ca}

\begin{abstract}
This paper contains a brief review of the remarkable properties of higher dimensional rotating black holes with the spherical topology of the horizon. We demonstrate that these properties are connected with and generated by a special geometrical object, the Principal Conformal Killing-Yano tensor (PCKYT). The most general solution, describing such black holes, Kerr-NUT-ADS metric, admits this structure. Moreover a solution of the Einstein Equations with (or without) a cosmological constant which possesses PCKYT is the Kerr-NUT-ADS metric. This object (PCKYT) is responsible for such remarkable properties of higher dimensional rotating black holes as: (i) complete integrability of geodesic equations and (ii) complete separation of variables of the important field equations.
\end{abstract}

\maketitle

\section{Introduction}

Main motivations for study higher dimensional black holes are nicely summarized in the paper by Harvey Reall. String theory, brane world models and ADS/CFT correspondence naturally involve higher dimensional gravity, and black holes play role of natural probes of extra dimensions.  Possibility of  creation of microscopic black holes in high energy colliders was (and still is) a subject which attracts a lot of intension. Besides these `rather technical' reasons there exists
another one of more general nature. The driving force of our scientific knowledge  is quite often our curiosity. We keep asking ourselves questions: "What happens if ...". From time to time by answering such questions one discovers interesting and non-trivial results.

\begin{figure}
  \begin{center}
  \includegraphics[width=8cm]{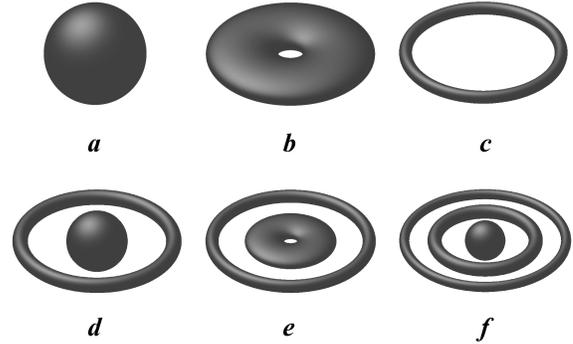}
  \end{center}
  \caption{\label{F1}Stationary 5D black holes.}
\end{figure}

By answering the question "Do black hole exist in higher dimensional gravity and what are their properties?" one understands better which of the properties of four-dimensional black holes are generic (valid in any number of dimensions) and which are specific in only our four-dimensional world. One of the most surprising discoveries was  that stationary black holes in higher dimensions can have large variety of  the horizon topology. The first of such 5-dimensional black objects with toroidal topology of the horizon $S^1\times S^2$, called black ring, was found by Emparan and Reall \cite{ER}. By now there are known many other stationary exact solutions describing vacuum black objects in 5D, the horizon of which is schematically shown at Figure~\ref{F1}. (For general discussion see the paper by Reall in this volume and a nice review \cite{LR}).

%{\footnotesize
%\begin{verbatim}
%\end{verbatim}}

Let us emphasize, that according to the general definition all such black objects,  strictly speaking, are {\em black holes}, since the region beyond their event horizon is causally separated from the future null infinity. However, in order to distinguish black objects with different horizon topology they received special nicknames (black ring, black Saturn ,etc), while the name of a black hole is reserved for the black object with spherical topology of the horizon. In this talk we shall follow this tradition.  This means that from now on any time when we  speak about a black hole we have in mind the object presented at the picture  1(a). To be more concrete
we focus our attention  on isolated higher dimensional rotating black holes with spherical topology of the horizon, obeying equations
\be\n{RR}
R_{\mu\nu}=\Lambda g_{\mu\nu}\, .
\ee
The corresponding spacetime is  asymptotically either  flat ($\Lambda=0$) or (anti)DeSitter one.

\section{Four dimensional Kerr-NUT-ADS metric and its higher dimensional generalization}

\subsection{Four dimensional Kerr-NUT-ADS spacetime}

The simplest example of a vacuum static black hole in 4D is the Schwarzschild metric
\be\n{mm}
ds^2=-f dt^2+dr^2/f +r^2 d\Omega_2^2\,  ,\quad f=1-r_0/r\, .
\ee
It contains only one parameter, the mass $M$ of the black hole, which determines the gravitational radius (size of the horizon) $r_0=2M$. Here and later we denote by $d\Omega_m^2$ the metric on a unit $m$-dimensional round sphere.

 A vacuum {\em stationary} black hole is characterized by one more parameter, $J$. Such a black hole  is rotating and $J$ is the value of its angular momentum. In fact the angular momentum (measured at infinity) is described by $3\times 3$ antisymmetric matrix $J_{ij}$. By rigid 3-dimensional rotations this matrix can be put in a standard form
\be\n{canJ}
{\bf J}=\left(\begin{array}{ccc}\displaystyle
0 & J & ~0~\\
-J & 0 & ~0~ \\
0 & 0 & ~0~
\end{array}\right)\, .
\ee
Kerr metric (with $J\le M^2$) is the most general vacuum stationary solution describing a regular black hole in an asymptotically flat spacetime.

In 1963 Newman, Tamburino and  Unti \cite{NUT} found another generalization of the Schwarzschild solution which besides mass $M$ contained another parameter $N$. This parameter $N$, called NUT-parameter, describes "gravitomagnetic monopole" \cite{Bonnor}. The corresponding NUT solution contains an analogue of the Dirac strings for a magnetic monopole, but in the gravitational case it cannot be excluded and does affect spacetime.  The most general stationary black hole solution of the Einstein equations (\ref{RR}) in an asymptotically (anti)deSitter spacetime  contains all these parameters $(M,J,N,\Lambda)$. It was obtained in \cite{Carter}  and it is called Kerr-NUT-ADS metric.

\subsection{Higher dimensional solutions}

Let us discuss now higher dimensional black hole solutions of the Einstein equations (\ref{RR}). We denote by $D$ the total number of (space and time) dimensions. Since some of relations have slightly different form in odd and even dimensions, we write
\be
D=2n+\ve\, .
\ee
It is quite easy to generalize the Schwarzschild solution to any number of spacetime dimensions. It is sufficient to substitute in (\ref{mm}) $d\Omega_{D-2}^2$ by the line element on $(D-2)-$ dimensional sphere $S^{D-2}$, and put $f=1-(r_0/r)^{D-3}$. This metric is known as Tangherlini solution \cite{Tang}.  The cosmological constant can also be easily included by using $f$ in the form
\be
f=1-(r_0/r)^{D-3}-{\Lambda\over D-1} r^2\, .
\ee

Before discussing higher dimensional rotating black holes let us make the following remark. In the asymptotically flat spacetime the total angular momentum of the objects, as measured at infinity, it described by an antisymmetric tensor $J_{ij}$, where $i$ and $j$ are spatial indices.
By suitable rigid rotations of the spatial coordinates, this $(D-1)\times (D-1)$-matrix can be transformed into the following canonical form:
\be
{\bf J}=\left( \begin{array}{ccccc}
0&J_{1}&0&0&\ldots \\
-J_{1}&0&0&0&\ldots \\
0&0&0&J_{2}&\ldots \\
0&0&-J_{2}&0&\ldots \\
\ldots &\ldots &\ldots &\ldots &\ldots
\end{array}
\right). \n{13.4.11}
\ee
It is easy to see that the total number of independent $2\times 2$ blocks is equal to $n-1+\ve$.
This means that there exist $n-1+\ve$ independent components of the angular momentum $J_i$, associated with  $n-1+\ve$ asymptotic independent two-dimensional spatial planes of rotation. The most general solution for a vacuum stationary rotating black hole was found in 1986 \cite{MP}. It contains $n+\ve$ independent constants: mass$M$  and components of the angular momentum $J_i$.

It took 20 years to discover the most general higher dimensional black hole solution, which besides mass and angular momentum contains also the cosmological constant and generalization of the NUT parameters \cite{KNAD}.
This solution is called a `general Kerr-NUT-ADS metric'. It is a natural generalization of the 4D Kerr-NUT-ADS metric. It contains $D-\ve$ arbitrary parameters, which include mass, $(n-1+\ve)$ rotation parameters, cosmological constant and $(n-1-\ve)$ NUT parameters.

\section{Principal Conformal Killing-Yano tensor}

\subsection{Definition}

The remarkable fact is: Properties of higher dimensional rotating black holes and their four dimensional `cousins'   are very similar. There exist a very deep geometrical reason for this similarity. All these metrics admit a special geometric object, the Principal Conformal Killing-Yano tensor (PCKYT), which is a generator of a complete set of explicit and hidden symmetries, that uniquely specifies the solution up to some constants. This solution coincides with Kerr-NUT-ADS metric.

 In fact this object is special case of a closed conformal Killing-Yano tensor. We shall give a general definition  of Killing-Yano tensors later.  We shall explain also  in more detail what are the hidden symmetries and how are they generated. At the moment, we focus on some remarkable properties of PCKYT.

Let us first give a definition of the PCKYT. Consider a 2-dimensional antisymmetric tensor (2-form) ${\bf h}$ which obeys the equation
\be\n{PCKY}
\nabla_c h_{ab}=g_{ca}\xi_b-g_{cb}\xi_a\, .
\ee
By contracting this equation one finds
\be\n{KV}
\xi_a={1\over D-1}\nabla^b h_{ba}\, .
\ee
If one antisymmetrizes the indices $a,b,c$ in (\ref{PCKY}), the right-hand side of this equation vanishes.This means that $\bf h$ is closed form, and (at least locally) can be presented in the  form
\be
\BM{h}=d\BM{b}\, ,
\ee
where $\bf b$ is a potential one-form.

Another important relation can be obtained if one takes the covariant derivative of (\ref{KV}) and symmetrises it. Using  an expression for commutator of covariant derivatives, and the Einstein equations(\ref{RR}) one proves that
\be
\xi_{(a;b)}=0\, .
\ee
In other words a spacetime admitting the 2-form $\bf h$ which is a solution of (\ref{PCKY}) always has a Killing vector. We call it a {\em primary Killing vector}.

We call $\bf h$ obeying (\ref{PCKY}) the Principal Conformal Killing-Yano tensor (PCKYT) if it is non-degenerate. More concretely this means the following:
 \begin{itemize}
 \item The matrix rank of $(D\times D)-$ antisymmetric matrix $h_{ab}$ is the largest possible, that is equal to $2n$.
 \item  Consider eigenvalue problem  for a matrix  $H^a_{\ b}=h^{a c}h_{c b}$
\be\n{CAN_COORD}
H^a_{\ b}e_{(i)}^b=x_ie_{(i)}^a\, .
\ee
It is easy to see that $h^a_{\ b} e_{(i)}^b$ is again an eigenvector with the same eigenvalue $x_{(i)}$. We assume that $\BM{H}$ has largest possible number, $n$, of different eigenvalues, and hence $n$ linearly independent eigen 2-planes.
\end{itemize}

\subsection{Remarkable properties of PCKYT}

The most general higher dimensional black hole metric, Kerr-NUT-ADS solution, admits the Principal Conformal Killing-Yano tensor. This result was first proved in \cite{FKL} for  Myers-Perry metrics, and later in \cite{KF} it was proved for the general Kerr-NUT-ADS spacetimes.

Moreover, a solution of the Einstein equations (\ref{RR}) in any number of dimensions ($D\ge 4$), which admits the Principal Conformal Killing-Yano tensor, is the Kerr-NUT-ADS spacetime. This result under special assumptions was first proved in \cite{HOY_1}. A general proof was given in \cite{KFK_UN,HOY_2}.

In other words, the existence of PCKYT is a characteristic property of  higher dimensional rotating black holes, with the spherical topology of horizon. Namely the existence of this object explains many remarkable properties of these spaces which sometimes are called `miraculous'. Before discussing this subject let us briefly remind  some properties of completely integrable dynamical systems.

\section{Complete Integrability}

\subsection{Liouville theorem}

Particle and light motion in a curved spacetime is described by geodesic equations. These equations are of the second order. Let $x^a(\tau)$ be a trajectory. By introducing a momentum  $p_a=g_{ab}\dot{x}^b$ as an independent variable, it is possible to rewrite the geodesic equations  in the first order form. These equations has the Hamiltonian form. This means that the general theory of dynamical systems can be applied for this problem. This approach is well known and its tools are very useful. Let us demonstrate this for the special problem: motion of a particle in a spacetime of a higher dimensional rotating black hole.

Consider a dynamical system It is described by an  even dimensional phase space $M^{2m}$ with a closed non-degenerate $2m-$symplectic form $\BM{\Omega}$ and a Hamiltonian $H$, which is a scalar function on the phase space. If $z^A$ are coordinates, then the dynamical equation is
\be
\dot{z}^A=\Omega^{AB}H_{,B}\, .
\ee
Poisson bracket for two functions $F$ and $G$  on the phase space is defined  as
\be
\{ F,G\}=-\Omega^{AB}F_{,A}G_{,B}\, .
\ee
These two functions are called to be in involution if their Poisson bracket vanishes.
A scalar function $F(z^A)$ on the phase space is
a first integral of motion if its Poisson bracket with the Hamiltonian vanishes $\{F,H\} = 0$ .

Liouville (1855) proved the following theorem: If a system with a Hamiltonian $H$ in the $2m$ dimensional  phase space has $m$ independent first integrals in involution, $F_1 =H, F_2, . . . , F_m$, then the system can be integrated by quadratures.
Such a system is called {\em completely integrable}.

\subsection{Relativistic particle as dynamical system}

Motion of a particle in a curved spacetime is special case of a dynamical system. If $D$ is the dimension of the spacetime and its coordinates are $x^a$, a particle trajectory is a line $x^a(\tau)$. The canonical coordinates in the corresponding phase space are $(x^a ,p_a=g_{ab}\dot{x}^b)$. The canonical sylleptic form and the Hamiltonian are
\be
{\BM{\Omega}}=dx^a \wedge dp_a\, ,
\quad H={1\over 2}g^{ab}p_a p_b\, .
\ee
It is easy to show that the Hamilton equations of motion in the phase space are equivalent to geodesic equations in the spacetime.

Let us assume that the Hamilton equations have an integral of motion
of the form ${\cal K}=K^{a_1\ldots a_q}(x) p_{a_1} \ldots p_{a_q}$ , then
\be
K_{(a_1\ldots a_q;a_{q+1})}=0\, .
 \ee
 Such a tensor in the spacetime is called a {\em Killing tensor}. The Killing tensor of the rank 1, $\xi_a$, is a Killing vector. The metric $g_{ab}$ is a trivial Killing tensor.

It is well known that Killing vectors generate symmetry transformation on the spacetime manifold with metric $g_{ab}$. Usually  this symmetry is called an {\em explicit symmetry}. The Killing tensors of higher rank more directly connected with the properties of the phase space, and their meaning is not so transparent in the spacetime itself. For this reason, they are often called {\em hidden symmetries}.

Poisson brackets in the phase space after reduction to the spacetime determine the {\em Schouten-Nijenhuis brackets}. When two integrals of motion are in the involution and each of them is a monomials in momentum, the  Schouten-Nijenhuis bracket for the corresponding Killing tensors vanish. In such a case we simply say that the Killing tensors commute.

The Liouville theorem being applied to relativistic particles implies that the geodesic equations are completely integrable (can be solved in quadratures) if there exist $D$ independent commuting Killing tensors. Some of them can be Killing vectors. $D$ dimensional Kerr-NUT-ADS admits $n+\ve$ Killing vectors. For complete integrability there must exist additionally $n$ Killing tensors. One of them is trivial, $g_{ab}$.

\subsection{Page's proposal}

Let $u^a=d{x}^a/d\tau$ is a unit tangent vector to a particle trajectory, then $P^{a}_b=\delta^{a}_b+u^a u_b$ is a projector to a plane, orthogonal to $u^a$. Using the definition of the PCKYT (equation (\ref{PCKY})) it is easy to show that $F_{ab}=P_a^c h_{cd} P^d_b$ is parallel transported along the geodesic. Hence any scalar invariant constructed from this tensor is a constant of motion. To obtain such scalars it is sufficient to take a trace of powers of the matrix $F_a^{\ b}$.  In \cite{PP} it was shown that these integrals of motion are independent and commuting, and their total number  is sufficient for the complete integrability of the geodesic equations in the Kerr-NUT-ADS spacetime. However these set of integral of motion is reducible. In other words most of these integrals are of high rank,while there exist a complete set of lower rank integrals of motion. In fact, in a general case in the presence of the PCKYT the irreducible set of integrals of motion is determined by $(n+\ve)$ Killing vectors and $n$ Killing tensors of the second rank. To demonstrate this we need to consider Killing and Killing-Yano tensors in more detail.

\section{PCKYT and Killing-Yano tower}

\subsection{Killing-Yano tensors}

Let us introduce first two objects. The {\em Killing-Yano (KY) tensor}
$k_{a_1\ldots a_q}$ is an antisymmetric $q-$form on the spacetime, which obeys the equation
\be
\nabla_a k_{a_1\ldots a_q}=\nabla_{[a}  k_{a_1\ldots a_q]}\, .
\ee
On the other hand, the {\em closed conformal Killing-Yano
(CCKY) tensor}  $h_{a_1\ldots a_q}$  is an antisymmetric $q-$form  the covariant derivative of which is determined
by its divergence
\be
\nabla_a h_{a_1\ldots a_q}=q g_{a[a_1}\xi_{a_2\ldots a_q]}\, ,
\ee
\be
\xi_{a_2\ldots a_q}={1\over D-r+1}\nabla_b h^b_{\ a_2\ldots a_q}\, .
\ee

KY and CCKY tensors are related to each other through
the Hodge duality: the Hodge dual of a KY form is a CCKY tensor, and vice versa. It is easy to check that if $k_{a_1\ldots a_q}$ is a Killing-Yano tensor, then
\be
K_{ab}=k_{a a_2\ldots a_q}k_b^{\ a_2\ldots a_q}\,
\ee
is a Killing tensor. We shall use the following schematic notation for this operation ${\bf K}={\bf k}\cdot{ \bf k}$.

CCKY tensors possess the following remarkable property: {\em An external product of two CCKY tensors is again a CCKY tensor.} This property was at first proved in the tensorial form in \cite{KKPF}. Slightly later a simple direct proof of this result was obtained in \cite{FFF} using the formalism of differential forms.

\subsection{Killing-Yano tower}

Let us return to our main `hero' -- Principal Conformal Killing-Yano tensor. This is a special case of CCKY tensor. Its additional properties are that it has tensor rank 2 and it is non-degenerate. Namely these properties make it so useful.  If the spacetime admits a PCKYT $\bf h$ it also has a whole set of other objects, which we call {\em Killing-Yano tower}. First of all one can define a set of external powers of $\bf $
\be
\BM{h}^{\wedge  (j)}=\underbrace{\BM{h}\wedge \ldots \wedge \BM{h}}_{j \mbox{   times}} \, .
\ee
All of these objects are CCKY tensors of different matrix rank, starting from 2 for $j=1$ till $2(n-1)$ for $j=n-1$.  Taking Hodge dual tensors for each of $\BM{h}^{\wedge  (j)}$ with $j=1,\ldots,n-1$ one obtains a set $n-1$ Killing-Yano tensors $\BM{k}^{(j)}$, and their squares
\be
{\bf K}^{(j)}={\bf k}^{(j)}\cdot{ \bf k}^{(j)}
\ee
determine $n-1$ Killing tensors of the rank 2.

For $j>n$ $\BM{h}^{\wedge  (j)}=0$. The case $j=n$ is special.
In the even dimensional case for $j=n$ one obtains an absolutely antisymmetric object. Its Hodge dual is a scalar. For the odd dimensional case and  $j=n$, the Hodge dual object is a Killing vector. One can also show that action of the Killing tensors ${\bf K}^{(j)}$ on the primary Killing vector gives a new independent (secondary) Killing vectors.

As a result of this construction starting with the PCKYT  $\bf h$ one obtains $(n-1)$ second rank Killing tensors and  $(n+\ve)$ Killing vectors. This, together with metric gives $D=2n+\ve$ integrals of motion. Additional check shows that they are independent and in involution. This proves that the geodesic motion in the spacetime with PCKYT (and hence in the most general Kerr-NUT-ADS metric) is completely integrable.

Complete integrability of physically interesting finite dimensional dynamical systems is quite rare property.  Examples of such systems include: motion in Euclidean space under central potential; motion in the two Newtonian fixed centers; geodesics on an ellipsoid (Jacobi, 1838); motion of a rigid body about a fixed point (Euler, Lagrange and Kowalevski); Neumann model (for more details and examples see e.g. the monograph \cite{BabelonEtal:2003}). Characteristic property of a completely integrable dynamical system is that its trajectories are regular and can be used to construct regular foliations of the phase space. The opposite case are  chaotic dynamical systems. It should be emphasized that the study of geodesic motion in higher dimensional black holes provides one with new wide class completely integrable dynamical systems, which might have interesting applications both in the theoretical and mathematical physics.

\section{Separation of variables}

\subsection{General remarks}

The complete integrability of geodesic equations in the higher dimensional black holes is closely related to the property of the {\em complete separation of variables} in some field equations in the same spacetime.

Consider a Hamiltonian $H({\bf p},{\bf q})$, where ${\bf p}=(p_1,\ldots,p_m)$ and ${\bf q}=(q^1,\ldots,q^m)$. The Hamilton-Jacobi equation for this Hamiltonian is a first order partial differential equation for function $S({\bf q})$ of the form
\be
H(S_{,{\bf q}},{\bf q})=0\, .
\ee
Here $S_{,{\bf q}}=(S_{,q^1}, \ldots, S_{,q^m})$. Suppose $q^1$ and $S_{,q^1}$ enter this equation only in a special combination $\Phi_1(S_{,q^1},q^1)$. Then the variable $q^1$ can be separated and a solution $S$ can be written in the form
\be
S=S_1(q^1,C_1)+S'(q^2,\ldots, q^m)\, , \quad \Phi_1(S_{,q^1},q^1)=C_1\, ,
\ee
and the new function $S'(q^2,\ldots, q^m)$ obeys a reduced Hamilton-Jacobi
\be
H_1(S'_{,q^2}, \ldots, q^2, \ldots )=0\, .
\ee
{\em Complete separation of variables} implies that the solution $S({\bf q})$ can be written in the form
\be
S=S_1(q^1,C_1)+S_2(q^2,C_1,C_2)+\ldots +S_m(q^m,C_1,\ldots, C_m)\, .
\ee
The constants $C_i$ generate first integrals on the phase space. When these integrals are independent and in involution, the system is integrable in the Liouville sense (see e.g \cite{Arnold}).

\subsection{Complete separation of variables in Kerr-NUT-ADS spacetime}

Complete integrability of geodesic equations in the four-dimensional Kerr metric was discovered by Carter\cite{carter:68} who succeeded to separate variables in the corresponding Hamilton-Jacobi equation. Similar approach does work also in five dimensional case for the Myers-Perry metric, written in a similar Boyer-Lindquist coordinates \cite{FS}. However numerous attempts to separate variable in the higher dimensional Myers-Perry metric were not successful, except some special cases when additional restrictions were imposed on the rotating parameters.

Nevertheless, the general Kerr-NUT-ADS spacetime with arbitrary number of dimensions allows complete separation of variables for main field equations. The reason is the following. The very property of the separation of variables implies that there exist such a special coordinate system, in which this property is valid. For example, the Boyer-Lindquist in the Myers-Perry spacetime with $D>5$ are not good for this purpose. The separation is possible if one makes additional restrictions of the parameters of the solution. The existence coordinates in which the separability takes place in the Kerr-NUT-ADS in a general case is connected with presence of the Principal Conformal Killing-Yano tensor.

Let us describe these special coordinates. We already mention that eigen-values $x_i$, determined by equation (\ref{CAN_COORD}), are independent, and they can be used as $n$ coordinates (at least in some spacetime domain). Moreover the spacetime with PCKYT has $(n+\ve)$ independent {\bf commuting} Killing vectors $\xi_{(j)}$, $j=0,\ldots, n-1+\ve$. Consider integral lines of these vector fields
\be
{dy^a\over d\psi_j}=\xi_{(j)}^a\, .
\ee
It can be shown that the set of  $(D=2n+\ve)-$quantities $(x_i,\psi_j)$ can be used as coordinates. Namely in these coordinates the complete separation of variables takes place. This was first demonstrated for Klein-Gordon and Hamilton-Jacobi equations in the higher dimensional Kerr-NUT-ADS in the paper \cite{FKK}. The massive Dirac equation has a similar property \cite{OY}. In fact these 3 different types of equations are closely related to on another. The Hamilton-Jacobi equation for  $S$ can be obtained as the eikonal equation in the lowest order of WKB approximation, by substituting $\varphi\sim \exp(iS)$ into the Klein-Gordon equation
\be
(\Box-m^2)\varphi=0\, .
\ee
On the other hand, the massive Dirac equation is just `a square root' of the Klein-Gordon equation.

\subsection{Complete integrability and separation of variables in weakly charged black holes}

Till now we discussed black hole solutions of the Einstein equations (\ref{RR}). It is interesting that their nice properties still remain valid for a wider sub-class of the black holes which are slightly charged. We assume that the cosmological constant vanishes, so that a Killing vector $\xi^a$  field obeys the equation
\be
\Box\xi_a=0\, .
\ee
This equation coincides with the Maxwell equation for the electromagnetic field potential $A_a$ in the Lorentz gauge $A^a_{\ ;a}=0$. Hence, one can consider $\xi^a$ as a test electromagnetic field on a given Kerr-NUT background and include the interaction of charged particles with it. For the primary Killing vector $\xi_{(0)}$ such a system describes a weakly charged black hole.  The secondary fields can be used to describe weakly magnetized one.

It is interesting that equations of motion of charged particles in  weakly charged higher dimensional black holes are completely integrable and the corresponding Hamilton-Jacobi and Klein-Gordon equations are completely separable \cite{FK}. Moreover, the complete separability also takes place for charged Dirac equations in a weakly charged black hole \cite{CFKK}.

In these notes I focused on the remarkable properties of higher dimensional rotating black holes with the spherical topology of the horizon. It was demonstrated that many of their properties are quite similar to the properties of their 4 dimensional Kerr-NUT-ADS `cousins'. The reason of this is the existence of the Principal Conformal Killing-Yano tensor, which determines quite `rigid' structure of the solutions and serves as a `seed' generating their explicit  and hidden symmetries. Many recent interesting generalization of this approach and its application is discussed in the review article of  Marco Cariglia included in this volume. It also contains plenty of important references. I need also to mention 3 general reviews on this subject \cite{rFK,rCKK,rYH} which contain a lot of additional information and references.

\section*{Acknowledgements}
The author thanks David Kubiznak for  useful remarks.
The author thanks the Natural Sciences and Engineering Research Council of Canada and the Killam Trust for the financial support. He also appreciate the financial support from CNRS and the hospitality of the University of Tours, where this paper was written.

\section*{References}
%\bibliography{ae100prg}

\begin{thebibliography}{21}


\bibitem{ER} R. Emparan  and H. S. Reall, Phys.Rev.Lett. 88 (2002) 101101

\bibitem{LR} R. Emparan and H. S. Reall,  Living Rev.Rel. 11 (2008) 6; e-Print: arXiv:0801.3471


\bibitem{NUT} E. Newman, L. Tamburino, and T. Unti, J. Math. Phys. 4 (1963)  915.

\bibitem{Bonnor} W. B. Bonnor. General Relativity and Gravitation, 24 (1992) 551



\bibitem{Carter} B. Carter,  Commun.Math.Phys. 10 (1968) 280; B Carter,  Physics Letters A,   26 (1968)  399

\bibitem{Tang} F. R. Tangherlini, Nuovo Cim. B108 (1993); 1253-1274, Erratum-ibid. B110 (1995) 1505

\bibitem{MP}R. C. Myers and M. J. Perry,  Ann. Phys. (N.Y.) 172  (1986)  304

\bibitem{KNAD} W. Chen, H. L\"{u}, and C. N. Pope, Class. Quantum Grav. 23   (2006)  5323


\bibitem{FKL} V. P. Frolov and D. Kubiznak, Phys. Rev. Lett.  98  (2007) 11101


\bibitem{KF} D. Kubiznak and V. P. Frolov,  Class.Quantum Grav. 24 (2007)  F1

\bibitem{HOY_1} T. Houri, T. Oota, and Y. Yasui,  Phys. Lett. B656 (2007)  214
\bibitem{KFK_UN} P. Krtous, V. P. Frolov, and D. Kubiznak,  Phys. Rev. D 78 (2008)   064022

\bibitem{HOY_2}	T. Houri, T. Oota, Y. Yasui, Class.Quant.Grav. 26 (2009) 045015

\bibitem{PP} D. N. Page, D. Kubiznak, M. Vasudevan and  P. Krtous, Phys.Rev.Lett. 98 (2007) 061102

\bibitem{KKPF} P. Krtous, D. Kubiznak, D. N. Page and V. P. Frolov JHEP 0702 (2007) 004

\bibitem{FFF} V. P. Frolov, Prog.Theor.Phys.Suppl. 172 (2008) 210-219

\bibitem{BabelonEtal:2003}     O. Babelon, D. Bernard and M. Talon  , Introduction to classical integrable systems, (Cambridge University Press), (2003)

\bibitem{Arnold} V. I. Arnold, Mathematical Methods of Classical Mechanics, Springer-Verlag (1989)

\bibitem{carter:68} B. Carter,  Phys.Rev. 174 (1968) 1559-1571

\bibitem{FS} Frolov V P and Stojkovic D, Phys. Rev. D 67  (2003) 084004;
Phys. Rev. D 68  (2003) 064011

\bibitem{FKK} Frolov V P, Krtous P and Kubiznak D, J. High Energy Phys. 02 ( 2007) 005

\bibitem{OY} T. Oota and Y. Yasui, Phys. Lett. B 659  (2008) 688

\bibitem{FK} V. P. Frolov, P.l Krtous, Phys.Rev. D83 (2011) 024016

\bibitem{CFKK} M. Cariglia, V. P. Frolov, P. Krtous, D. Kubiznak, "Electron in higher-dimensional weakly charged rotating black hole spacetimes", arXiv:1210.????

\bibitem{rFK} V. P. Frolov and D. Kubiznak, Class.Quant.Grav. 25 (2008) 154005

\bibitem{rCKK} M. Cariglia, P. Krtous, D. Kubiznak, Fortsch.Phys. 60 (2012) 947-951

\bibitem{rYH} Y. Yasui and T. Houri,  Prog.Theor.Phys.Suppl. 189 (2011) 126-164


\end{thebibliography}

\end{document}